\documentclass[pra, twocolumn, showpacs, preprintnumbers, amsmath, amssymb, superscriptaddress, aps]{revtex4}
\usepackage{amssymb}
\usepackage{latexsym}

\usepackage{epsfig}

\usepackage{dcolumn}
\usepackage{amsmath}
\usepackage{amsfonts}
\usepackage{bm}

\newcommand{\be}{\begin{equation}}
\newcommand{\ee}{\end{equation}}
\newcommand{\bea}{\begin{eqnarray}}
\newcommand{\eea}{\end{eqnarray}}

\newcommand{\p}{\partial}
\newcommand{\s}{\sigma}

\newcommand{\rd}{\mbox{d}}
\newcommand{\ri}{\mbox{i}}

\begin{document}
\title{Majorana fermion realization of 2-channel Kondo effect in a junction of three quantum Ising chains}
\author{A. M. Tsvelik}
\affiliation{Department of Condensed Matter Physics and Materials Science, Brookhaven National Laboratory, Upton, NY 11973-5000, USA}
 \date{\today } \begin{abstract} 
It is shown that a junction of three quantum Ising chains ($\Delta$-junction) can be described as the 2-channel Kondo model in a box which size is of the order of the Ising model correlation length with spin S=1/2 localized at the junction. The local spin is composed of the zero energy boundary Majorana modes of the Ising models. 
\end{abstract}


\maketitle

Majorana fermions in general and Majorana zero energy modes (MZEMs) in particular have become a very popular topic. The interest is mainly caused by non-Abelian properties of MZEMs related to the fact that for a set of $N$ modes the corresponding operators satisfy the Clifford algebra 
\be
\{a_i,a_j\} = \delta_{ij}, ~~i,j=1,...N, \label{Cliff}
\ee
and as such represent a realization of a spinor representation of the O(N) group.

 Theory predicts that MZEMs will appear at topological defects such as vortices in two- and domain walls in one dimension.  Majorana fermion $a$ is a linear combination of fermionic creation and annihilation operator 
\be
a = (f +f^+)/2, ~~a^2= 1/2, ~~ a= a^+,
\ee
and as such cannot be used to create or destroy any state. To do this one needs a pair of Majorana fermions $a_1,a_2$ from which creation and annihilation operators can be constructed:
\be
f=a_1 +\ri a_2, ~~ f^+ = a_1 -\ri a_2.
\ee
In systems with periodic boundary conditions Majorana modes appear in pairs; their tunneling Hamiltonian is
\be
H_{tunn} = 2\ri t a_1a_2 = t(f^+f -1/2).
\ee
Thus for a finite chain the ground state degeneracy is broken, but since $t$ decays fast with the system size, for sufficiently large systems one can think about boundary Majorana modes as independent. Due to their nonlocality such Majorana modes have attracted a particular attention  in conjunction with quantum computation.  Although in one dimension the task of generating Majorana modes is much easier, it is more difficult  to take advantage of the non-Abelian nature of MZEMs since to exchange them  one needs somehow to escape from one dimension. It has been suggested that one can resolve the conundrum with a help of the so-called $Y$- or $T$-junction.  In \cite{oreg} it was suggested to use  $T$-junctions  to implement  braiding operations of Majorana fermions. 

 In this paper I study a slightly more complicated junction  of three quantum Ising models coupled together at one point, called а $\Delta$-junction. 
The Hamiltonian  describes three quantum Ising chains 
\bea
H_{Ising,p} = \sum_{j=1}^N\Big[-J\s^x_p(j)\s^x_p(j+1)  + h\s^z_p(j)\Big] \label{Ising}
\eea
coupled together in the $\Delta$-junction:
\bea
&& H = -J_{12}\s^x_1(1)\s^x_2(1) - J_{23}\s^x_2(1)\s^x_3(1) - J_{13}\s^x_1(1)\s^x_3(1) \nonumber\\
&& + \sum_{p=1}^3H_{Ising,p} \label{Y}
\eea
I assume that $0< J_{pq} << J$. $\Delta$-junction becomes $T$-junction when one of the exchange integrals $J_{ab}$ is zero. Models (\ref{Ising}) are equivalent to noninteracting Majorana fermions by means of Jordan-Wigner transformation. To extend this formalism for Ising models on star graphs Crampe and Trombettoni suggested to introduce  Klein factors \cite{crampe} to mark different chains:
 \bea
&& c_p(j) = a_p\left(\prod_{k=1}^{j-1}\s^z_p(k)\right)\s^-_p(j), \label{ferm}\\
&& \s^-_p(j) = a_p c_p(j)\exp\left[\ri\pi\sum_{k=1}^{j-1}c^+_p(k)c_p(k)\right], \nonumber\\
&& \s^z_p(j) = c^+_p(j)c_p(j)- 1/2, \nonumber
\eea
where  Klein factors $a_p$ satisfy the Clifford algebra (\ref{Cliff}) and commute with all  spin operators. Anticommutativity of $a$'s from different chains establishes  commutativity of the spin operators located at  different chains. Hence the emergence of  MZEMs in the fermionic formulation of the problem is quite natural.  Notice also that the boundary spins a have particularly simple expression in terms of fermions:
\be
\s^x_p(1) = a_p[c_p(1)- c^+_p(1)],  \label{boundary}
\ee
which is a lattice generalization of the expression obtained in \cite{goshal} (see Eq. (4.25)). 

Substituting (\ref{ferm}) into (\ref{Ising},\ref{Y}) we obtain the following fermionic Hamiltonian:
\bea
&& H = \sum_{p=1}^3 H_{Ising,p} +V\label{Hferm}\\
&& H_{Ising} = \sum_{j=1}\Big\{-J[c^+(j+1)-c(j+1)][c^+(j)+c(j)] + \nonumber\\
&& hc^+(j)c(j)\Big\}\label{Isferm}\\
&& V =\sum_{p>q} J_{pq}a_pa_q[c_p(1)-c^+_p(1)][c_q(1)-c^+_q(1)] \label{Vferm}
\eea
As we see, the MZEMs represented by the Klein factors do not appear in the expressions for the bulk Hamiltonians (\ref{Isferm}). They appear only in the term describing the junction and  for this reason their combination can be interpreted as a quantum degree of freedom located at the junction. For a junction of three chains this degree of freedom can be described as S=1/2 spin \cite{martin}:
\be
S^p = \frac{\ri}{2}\epsilon_{pqt}a_qa_t, \label{BS}
\ee

 It will be shown that for nearly critical chains $|J-h| << J$ the MZEMs  are subject of intense screening by  gapless bulk modes and  model (\ref{Isferm},\ref{Hferm}) is equivalent to the overscreened two-channel Kondo model. The latter model is interesting since it possesses non-Fermi liquid fixed point \cite{blandin} with a residual ground state entropy \cite{wiegmann},\cite{tsvelik}. The two-channel Kondo model has been extensively studied and a plethora of nonperturbative results have been obtained for it by means of Bethe ansatz \cite{andrei},\cite{wiegmann},\cite{tsvelik} and conformal field theory (see, for instance, \cite{AL}). The equivalency between the $\Delta$-junction model (\ref{Y}) and the two-channel Kondo model allows one to use these results. 


 To establish the aforementioned relation between the junction of critical Ising chains and the Kondo model we need to switch to the continuous description of model (\ref{Isferm},\ref{Hferm}) valid for $|J-h| << J$. For this end we use the Majorana fermions 
\bea
\xi(j )= c^+(j) + c(j), ~~ \rho(j) = \ri[c^+(j)-c(j)],
\eea
pass to the continuum limit and introduce the right- and the left-moving modes $\chi_{R,L}(x) = [\rho(x=jb) \pm\xi(x=jb)]/\sqrt {2b}$, where $b$ is the lattice constant. The resulting Lagrangian for (\ref{Isferm}) is 
\bea
 && L = \int_{0}^{\infty}\rd x \Big[\frac{1}{2}\chi_R(\p_{\tau} - \ri v\p_x)\chi_R + \frac{1}{2}\chi_L(\p_{\tau} + \ri v\p_x)\chi_L + \nonumber\\
&& \ri M\chi_L\chi_R\Big] + \frac{\ri}{2}a\p_{\tau}a,\label{cont}
\eea
where $v = Jb, ~~ M = J-h$ and a free boundary condition 
\bea
\chi_R(0) = \chi_L(0) \label{free}
\eea
This condition corresponds to the fact that the boundary spin (\ref{boundary}) is not fixed. Lagrangian (\ref{cont}) coincides with the one obtained for the Ising model with the boundary in \cite{goshal}. Let us now set $M=0$. With boundary condition (\ref{free}) one can introduce a chiral fermion $\chi(x) = \theta(x)\chi_R(x) +\theta(-x) \chi_L(-x)$ and extend the integration in (\ref{cont}) on the entire $x$-axis.  This can be done safely since $\chi(x)$ is continuous at $x=0$. Taking into account  interaction term (\ref{Vferm})  and switching back to the Hamiltonian formalism, I arrive to the following effective theory for the $\Delta$-junction of three critical Ising chains:
\bea
 H_{eff} = \ri\sum_p G_pS^p\epsilon_{pqt}\chi_q(0)\chi_t(0) + \frac{\ri v}{2}\int_{-\infty}^{\infty} \rd x \chi_p\p_x\chi_p , \label{Kondo}
\eea
where $G_1 = bJ_{23}, G_2 = bJ_{13}, G_3=b J_{12}$. This description is valid at energies $<< J$. I used the equivalence between bilinears of MZEMs and components of spin S=1/2 (\ref{BS}) to replace them with the spin operator. 

  Model (\ref{Kondo}) describes the two-channel Kondo problem written in the  form  introduced in \cite{ioffe}. This equivalence is based on the fact  that the fermionic bilinears coupled to the "spin" 
\[
{\cal J}^a = \frac{\ri}{2}\epsilon_{abc}\chi_b\chi_c, 
\]
are SU$_2$(2) currents, that is satisfy the same commutation relations as the corresponding fermionic bilinears in the 2-channel Kondo model. 

 If the bare interactions are ferromagnetic ($-J_{ab}< 0$, the same sign as the bulk exchange), the Kondo exchange in (\ref{Kondo})  is antiferromagnetic and the interaction scales to the intermediate coupling critical point. Otherwise it is marginally irrelevant. A similar Kondo model (the 4-channel one) has recently been suggested in \cite{cooper} in the context of the so-called topological Kondo effect. In \cite{crampe} it was found that 4-channel Kondo model describes $\Delta$-junction of XX spin S=1/2 chains. 

As was established in \cite{ioffe}, the low energy Lagrangian describing the junction dynamics at energies less than the Kondo temperature $T_K$:
\bea
L_{eff} = \frac{\ri}{2}\epsilon\p_{\tau}\epsilon + g\epsilon\chi_1(0)\chi_2(0)\chi_3(0) + \sum_{p=1}^3L[\chi_p], 
\eea
where $g\sim T_K$ and $L[\chi_p]$ describes three chiral Majorana modes of the bulk and $\epsilon$ is a new MZEM describing the residual degeneracy of the ground state. This MZEM is nonlocal in terms of the fields of model (\ref{Kondo}). The critical point is characterized by a single zero energy Majorana fermion coupled to the bulk by the irrelevant operator.

The scaling towards this boundary critical point takes place even if one of the couplings is zero.  Then the $\Delta$-junction becomes  a $T$-junction. Indeed, the first loop renormalization group equations for the running coupling constants are
\bea
\frac{\rd g_a}{\rd \ln\Lambda} = - g_bg_c, ~~ (b \neq c \neq a),
\eea
where $g_a(J) = 2G_a/\pi v$. Hence even if, for instance, $g_1(J) =0$, the coupling $g_1$ will be generated by two other couplings under renormalization. Thus the model flows to the 2-channel Kondo critical point  irrespectively of the ratios between the couplings $G_a$; it is well known that the multichannel Kondo model fixed  point is unaffected by the anisotropy \cite{pang}.  This is a  remarkable fact meaning that  the low energy properties of the junction are  robust with respect to anisotropy of the couplings.The Kondo scale $T_K$ corresponds to the energy when all dimensionless coupligs $g_a$  become $\sim 1$ and is a function of the bare couplings $G_a$ and the ultraviolet cut-off $J$, for equal couplings $G_1 = G_2 = G_3$ it is exponentially small in $G$.  

 Another remarkable fact is that the local pseudospin (\ref{BS}) constructed from the MZEMs of different chains is not a local operator of Ising chains and therefore there is no "magnetic field" which can be attached to it.   However, the critical flow can be destroyed by application of an additional  boundary magnetic field $\delta h$ on at least one chain (for instance, to the chain number 1). According to (\ref{boundary}) this introduces a relevant operator
\bea
\delta H = \ri \delta h a_1\chi_1(0).
\eea
This perturbation effectively eliminates the terms with $G_2,G_3$ in the interaction term in  (\ref{Kondo}). The remaining term contains just one spin component and as a result does not experience any renormalization. This is equivalent to decoupling of one of the chains from the junction.


 Let us consider two $\Delta$-junctions coupled together. Zero modes from one of them we denote $\eta^-$, from the other $\eta^+$. The Hamiltonian (\ref{Ising},\ref{Y}) acquires a correction in the form of tunneling term:
\be
\ri t_a\eta^+_a\eta^-_a, ~~ t\sim 1/L.
\ee
This term competes with the Kondo screening which can be established looking at the mean field solution where the interaction is decoupled by fields $\Delta_{\s}$ so that the Lagrangian becomes (I consider the isotropic interaction):
\bea
L[\chi^{\s}] + \Delta_{\s}^2/2G + \ri\Delta_{\s}\eta_a^{\s}\chi_a^{\s}(0) + \ri t\eta_a^+\eta_a^-, ~~\s = \pm 1.
\eea
The saddle point action is 
\bea
\frac{1}{2G}(\Delta_+^2 + \Delta_-^2) - \frac{3}{8\pi}\int \rd\omega \ln[(|\omega|+\Delta_+)(|\omega| +\Delta_-)+ t^2]
\eea
yielding 
\be
\Delta_{\s} = T_K - t, ~~ (t < T_K); ~~ \Delta_{\s} =0, ~~ (t > T_K)
\ee
where $T_K \sim \exp(- Const/G)$. From here we may conclude that  the Kondo effect disappears at $t>T_K$. 

 The current study indicates that a $\Delta$-junction (with $T$-junction being its particular case)  of nearly critical quantum Ising models behaves as an active element where MZEMs undergo screening by the bulk excitations. At low energies the system scales to overscreened 2-channel Kondo model critical point. This is an interesting fact by itself since realization of overscreened Kondo effect is a notoriously difficult task. At the same time nearly critical quantum Ising systems can be realized in Josephson junction arrays \cite{gersh}.

 I am  grateful to A. A. Nersesyan for bringing this problem to my attention and extended subsequent discussions and to B. L. Altshuler and A. B. Zamolodchikov for the interest to the work. I am also grateful to Artem Abanov for pointing out a difference between $\Delta$ and $T$-junctions. The work  was  supported  by US DOE under contract number DE-AC02 -98 CH 10886.

\end{document}